\documentclass[11pt,a4paper]{article}
\usepackage{jheppub}
\usepackage[english]{babel}
\usepackage{amssymb,amsmath}
\usepackage{mathrsfs}
\usepackage{hyperref}
\usepackage{color}
\usepackage{graphicx}
\usepackage{mathtools}
\usepackage{subfigure}
\usepackage{slashed}
\usepackage{float}
\usepackage{booktabs}

\DeclarePairedDelimiter\abs{\lvert}{\rvert}%
\newcommand{\zir}{{z_{\rm IR}}}

\title{Holographic Energy Correlators for Confining Theories}

\author{Csaba Cs\'aki}
\author{and Ameen Ismail}
 
\affiliation{Laboratory for Elementary Particle Physics, Cornell University, Ithaca, NY 14853, USA}

\emailAdd{csaki@cornell.edu}
\emailAdd{ai279@cornell.edu}

\abstract{We present a holographic calculation of energy correlators in a simple model of confinement based on a warped extra dimension with an IR brane. For small distances we reproduce the constant correlators of a strongly-coupled conformal field theory, while for large distances the effects of confinement dominate and the correlators decay exponentially. We find exact shockwave solutions to the Einstein equations in the presence of the IR brane, hence avoiding the need for a perturbative expansion in terms of Witten diagrams. While some of the expected qualitative features of energy correlators in quantum chromodynamics (QCD) are reproduced, our crude model of confinement does not capture the effects of asymptotic freedom nor exhibit jetty behavior. We expect that our method can also be applied to more realistic models of confinement incorporating asymptotic freedom, which should fix some of the deviations from QCD. 
}

\begin{document}

\maketitle	
\flushbottom

%%%%%%%%%%%%%%%%%%%%%%%%%%%%%%%%%%%%%%%%%%%%%%%%%%%%%%%%%%%
%%%%%%%%%%%%%%%%%%%%%%%%%%%%%%%%%%%%%%%%%%%%%%%%%%%%%%%%%%%
\section{Introduction}
%%%%%%%%%%%%%%%%%%%%%%%%%%%%%%%%%%%%%%%%%%%%%%%%%%%%%%%%%%%
%%%%%%%%%%%%%%%%%%%%%%%%%%%%%%%%%%%%%%%%%%%%%%%%%%%%%%%%%%%

Understanding the dynamics of confinement in quantum chromodynamics (QCD) is one of the last important unsolved problems of the Standard Model (SM) of particle physics. QCD becomes strongly coupled around the characteristic scale $\Lambda_{\rm QCD} \sim   200$~MeV, below which it is better described as a theory of composite hadrons rather than that of quarks. Due to this strong coupling it is very difficult to get a firm theoretical handle on QCD from first principles. One is forced to consider numerical studies via lattice simulations, consider various models of QCD, or study supersymmetric cousins of QCD to gain some qualitative understanding. 

One area of rapid recent progress in understanding the dynamics of QCD has come through the study of energy correlators. These are correlation functions of the energy flow operators $\mathcal{E}(\vec{n})$ of the sort $\langle \mathcal{E}(\vec{n})\rangle $, $\langle \mathcal{E}(\vec{n}) \mathcal{E}(\vec{n}')\rangle $, etc.  The physical meaning of the energy flow operators in a collider experiment is to measure the energy deposited in a calorimeter placed very far away from the interaction point, in the direction specified by the unit vector $\vec{n}$.

Energy correlators were first considered in~\cite{Basham:1977iq,Basham:1978bw,Basham:1978zq,Basham:1979gh}, but their study was reinvigorated by Hofman and Maldacena~\cite{Hofman:2008ar,Hofman:2009ug}, who discussed energy correlators in conformal field theories (CFTs). They also pointed out that for CFTs that admit a gravitational dual as in the AdS/CFT correspondence~\cite{Maldacena:1997re,Witten:1998qj,Gubser:1998bc}, energy correlators can be calculated holographically in the dual theory~\cite{Hofman:2008ar}. Investigations of energy flow and related light-ray operators have led to progress in our understanding of CFT and quantum field theory (QFT) more generally, for example~\cite{Zhiboedov:2013opa,Belitsky:2013xxa,Belitsky:2013bja,Bousso:2015wca,Faulkner:2016mzt,Bousso:2016vlt,Hartman:2016lgu,Casini:2017roe,Balakrishnan:2017bjg,Cordova:2017zej,Cordova:2017dhq,Leichenauer:2018obf,Kravchuk:2018htv,Cordova:2018ygx,Ceyhan:2018zfg,Manenti:2019kbl,Balakrishnan:2019gxl,Korchemsky:2021htm,Hartman:2023qdn} (for further references see~\cite{Lee:2022ige}).

Over the past five years, there has been a very large and highly successful effort to connect formal developments in conformal collider physics to real-world phenomenology~\cite{Dixon:2019uzg,Chen:2019bpb,Chen:2020vvp,Chen:2020adz,Chen:2021gdk,Komiske:2022enw,Holguin:2022epo,Chen:2022jhb,Chen:2022swd,Lee:2022ige}. It is now possible to extract energy correlators from Large Hadron Collider (LHC) data, offering new insights into the dynamics of confinement~\cite{Komiske:2022enw,CMS:2024mlf}. The two-point correlator in particular provides a very clean view of the confinement transition.
The degree of angular separation between the two energy flow operators corresponds to the transverse momentum. Thus at small angular scales the correlator probes low-energy, long-distance behavior, scaling as a power law characteristic of free hadrons. At larger scales one observes a nearly constant correlator, with a logarithmic running corresponding to asymptotically free quarks. In this regime one can even observe the correct scaling behavior predicted by perturbative QCD~\cite{Lee:2022ige}.

Despite these important advances, reproducing the structure of the confinement transition in the two-point correlator from first principles remains elusive, even at a qualitative level. In the confinement transition regime neither perturbative QCD techniques nor the free hadron description are applicable for evaluating energy correlators. A reasonable approach is then to try using a model of QCD to find the behavior of energy correlators in this regime. The most plausible approach would be to use AdS/QCD; that is, to model the dynamics of QCD with a 5D gravitational dual (for examples and reviews see~\cite{Sakai:2004cn,deTeramond:2005su,Erlich:2005qh,DaRold:2005mxj,Sakai:2005yt,Karch:2006pv,Csaki:2006ji,Csaki:2008dt,Erlich:2009me}). The formal developments in computing energy correlators from AdS/CFT, together with the abundance of holographic models of QCD, beg for a holographic treatment of the confinement transition observed in the two-point correlator.

In this paper we take the first step toward a holographic computation of QCD energy correlators by calculating the two-point correlator in the simplest 5D model of confinement, corresponding to pure AdS cut off by an infrared (IR) brane~\cite{Randall:1999ee}. Of course, this does not capture all of the dynamics of energy correlators in QCD.
In particular the high-energy behavior corresponds to that of a strongly-coupled CFT rather than asymptotically free quarks.
Despite these shortcomings, we observe a clear transition between the confined and deconfined regimes. This is the first computation of energy correlators in a confining holographic model. We hope that in studying a heavily simplified model of confinement, we lay the groundwork for future calculations of energy correlators in more realistic AdS/QCD models.

We take a similar approach to~\cite{Hofman:2008ar,Hofman:2009ug,Belin:2020lsr} in calculating energy correlators. The basic idea is to study shockwave solutions, in which the AdS metric is perturbed by a source localized in the lightcone coordinate $x^+$; in the 4D picture, this corresponds to the insertion of an energy flow operator in the path integral. The profile of shockwaves about the full AdS spacetime is well-known. Our main achievement is to find a closed form for the shockwave when we cut off AdS with an IR brane. Importantly, arbitrary linear superpositions of these shockwaves are exact solutions to the Einstein equations. Because of this, one can superpose $n$ shocks to compute $n$-point energy correlators. This conveniently avoids the need to expand the action in a power series in the gravitational coupling and evaluate Witten diagrams.

This paper is organized as follows. We first review the definition of energy-energy correlators and how they are computed in AdS/CFT in Section~\ref{sec:review}. The holographic calculation essentially relies on the study of the shockwave geometries mentioned previously. In Section~\ref{sec:IRbrane} we cut off AdS with an IR brane and compute the two-point correlator with a scalar source. The presence of the IR brane modifies the boundary conditions for the shockwave, thereby altering the geometry and ultimately the form of the correlator. We present numerical results for the two-point correlator and derive analytical expressions for its asymptotic behavior in the small-distance and large-distance limits. At length scales smaller than the IR brane location we obtain a nearly constant correlator, characteristic of a strongly-coupled CFT.
At distances larger than the IR brane location the correlator decays exponentially, corresponding to confinement.
In Section~\ref{sec:discussion} we conclude and comment on future directions for energy correlators, holography, and confinement.

%%%%%%%%%%%%%%%%%%%%%%%%%%%%%%%%%%%%%%%%%%%%%%%%%%%%%%%%%%%
%%%%%%%%%%%%%%%%%%%%%%%%%%%%%%%%%%%%%%%%%%%%%%%%%%%%%%%%%%%
\section{Holography for conformal colliders}\label{sec:review}
%%%%%%%%%%%%%%%%%%%%%%%%%%%%%%%%%%%%%%%%%%%%%%%%%%%%%%%%%%%
%%%%%%%%%%%%%%%%%%%%%%%%%%%%%%%%%%%%%%%%%%%%%%%%%%%%%%%%%%%

In this section we review the holographic computation of energy correlators following~\cite{Hofman:2008ar,Belin:2020lsr}. We imagine some excitation of a CFT localized at the origin $x^\mu = 0$, produced by a perturbation external to the CFT. In a phenomenological context, this would correspond to the production of CFT stuff from e.g. an $e^+e^-$ collision. This $e^+e^-$ collision will be localized on the AdS boundary (a ``UV brane", though the existence of a UV brane is not essential here and in most cases we will assume the UV brane is moved all the way to the AdS boundary)  and will generate some CFT radiation which will propagate into the bulk of the AdS. A simple way to represent the specific CFT perturbation excited by the collision is to introduce a bulk field corresponding to it, which will couple to the source of the excitation (the $e^+e^-$ collision) on the UV brane. The simplest possibility is to consider a bulk scalar field $\phi$ , but this is just one of many possible choices. This bulk scalar will start falling into the bulk of the AdS, but eventually the corresponding CFT radiation will be detected by some calorimeters far away from the source. The calorimeter is made of ordinary matter, hence it should be localized on the UV brane far out from the source. The calorimeter measures the energy deposited, which corresponds to the appropriate component of the energy-momentum tensor. Thus we can interpret this measurement as looking for a correlator involving a certain number of insertions of the energy-momentum tensor.

\begin{figure}
    \centering
    \includegraphics[width=0.3\textwidth]{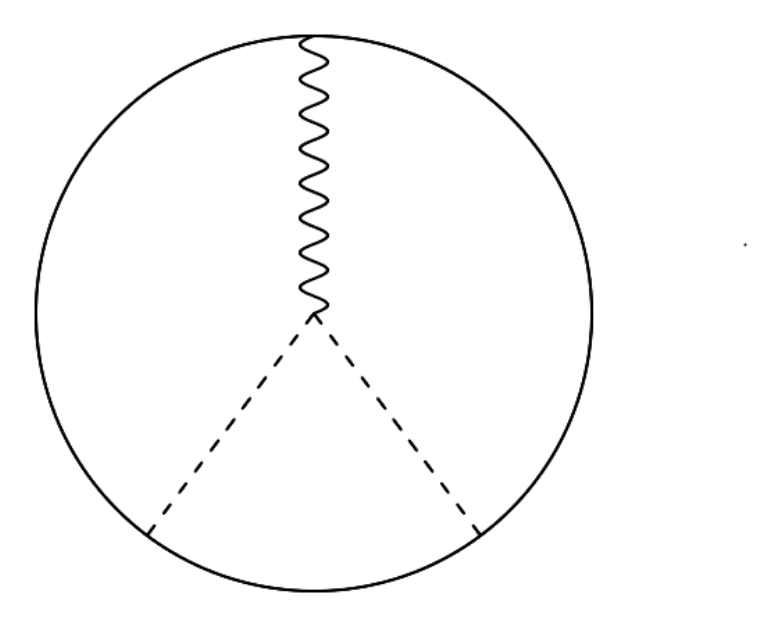}\hspace*{0.5cm}
    \includegraphics[width=0.245\textwidth]{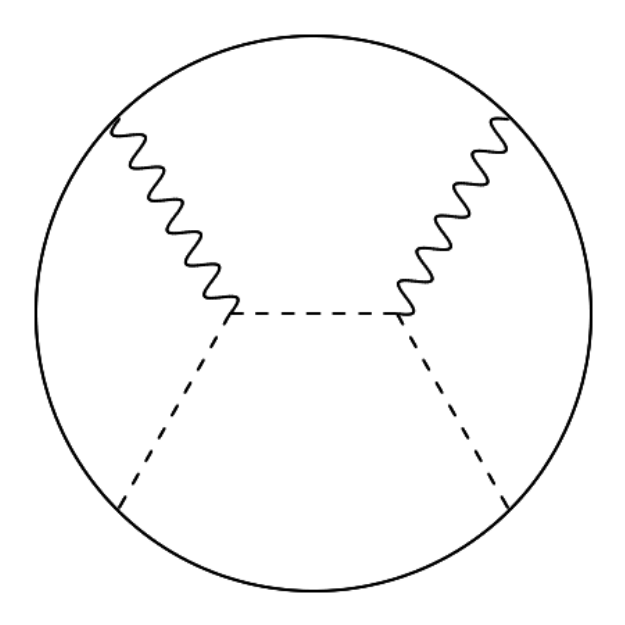}
    \includegraphics[width=0.245\textwidth]{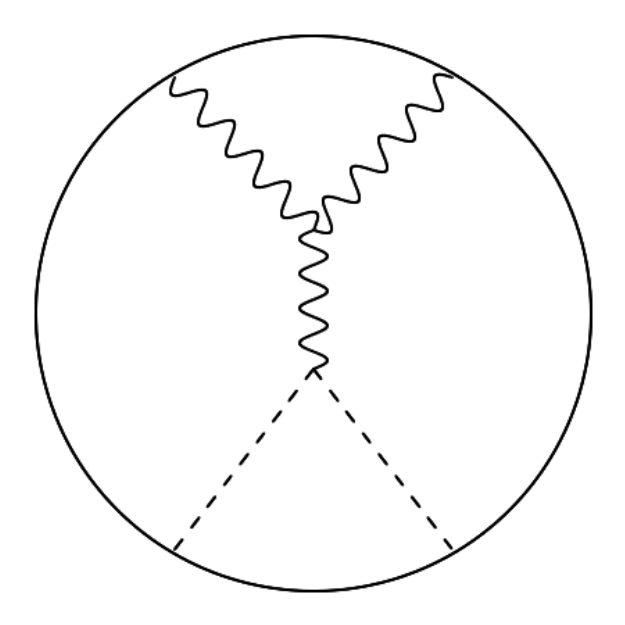}
    \caption{Left: The leading Witten diagram contributing to the one-point energy correlator. Right: The leading Witten diagrams contributing to the two-point correlator.}
    \label{Fig:Witten1}
\end{figure}

\begin{figure}
    \centering
    \includegraphics[width=0.245\textwidth]{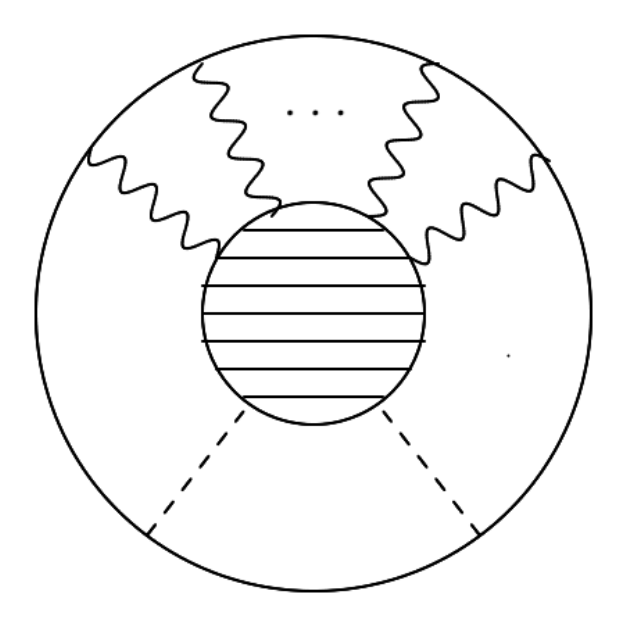} 
    \caption{Sketch of the Witten diagram for the general $n$-point energy correlator. The method of shockwaves circumvents the calculation of these diagrams and can be used for any $n$.}
    \label{Fig:Witten2}
\end{figure}

On the CFT side the insertion of the energy-momentum tensor can be interpreted as a source for bulk gravitons. Hence, the holographic calculation of an $n$-point energy correlator corresponds to a calculation of an $n+2$-point function with two insertions of the external source (creating the bulk scalar fields) and $n$ insertions of the energy-momentum tensor creating $n$ bulk gravitons. In the AdS/CFT language this is represented by so-called Witten diagrams, where bulk scalars emanating from the UV boundary propagate back to the boundary in the form of bulk gravitons. The corresponding Witten diagram for the one- and two-point functions are shown in Fig.~\ref{Fig:Witten1}, while a generic $n$-point function is sketched in Fig.~\ref{Fig:Witten2}. 

AdS/CFT provides a systematic prescription for evaluating these Witten diagrams, however they are technically quite involved. Hofman and Maldacena~\cite{Hofman:2008ar} presented an alternative holographic calculation based on the method of shockwaves. They found that the Einstein equation corresponding to the insertion of the energy-momentum tensor on the boundary can actually be exactly solved by some simple shockwave geometries. An energy insertion will then generate one of these shockwaves. One can also consider the superposition of several of them, which remain exact solutions to the Einstein equations. One can then find energy correlators by considering the intersection of the shockwaves with the propagating bulk fields excited by the external source on the boundary. Below we will review the details of the Hofman--Maldacena calculation, setting up our generalization to theories with a mass gap in the next section. 

\subsection{Energy correlators}

The basic objects for the calculation of energy correlators are the energy flow operators defined as 
\begin{equation}\label{eq:energyflow}
    \mathcal{E}(\vec{n}) = \lim_{r \rightarrow \infty} r^2 \int_{0}^\infty dt\, T_{0i}(t, x^i = r n^i)\, n^i ,
\end{equation}
where $\vec{n}$ is a unit vector with components
\begin{equation}
    n^i = (\sin \theta \cos \phi, \sin \theta \sin \phi, \cos \theta)\ .
\end{equation}
We reiterate that these operators measure the energy deposited in a calorimeter placed very far away from the interaction point. As written the limit in Eq.~\eqref{eq:energyflow} is ambiguous. For a conformal theory we expect massless excitations, hence we expect energy to propagate out to future null infinity. Thus we should take the large $r$ limit while holding $t - r$ constant and sending $t + r \rightarrow \infty$~\cite{Belitsky:2013xxa,Chen:2019bpb}.

We want to consider correlation functions of the energy flow operators, such as a two-point correlator
\begin{equation}
    \langle \mathcal{E}(\vec{n}) \mathcal{E}(\vec{n}') \rangle .
\end{equation}
Here the expectation value is evaluated on the state created by the external perturbation.

For calculations, it is useful to express the energy flow operators, Eq.~\eqref{eq:energyflow} in terms of lightcone coordinates $x^\pm = x^0 \pm x^3$. The energy flow operator then takes the form
\begin{equation}\label{eq:energyflow2}
    \mathcal{E}(\vec{n}) = \lim_{x^+ \rightarrow \infty} \frac{\left( x^+ \right)^2}{4} \int_{-\infty}^\infty dx^- T_{--}(x^+, x^-, x^\perp) ,
\end{equation}
where $x^\perp = x^{1,2}$.
One nice property of this expression is that it makes it manifest that these operators are integrals over the future null boundary of Minkowski space. In terms of $t$ and $r$ the lightcone coordinates can be expressed as
\begin{equation}\label{eq:coordinates}
    x^\pm = (t - r) + r \left(1 \pm \cos \theta \right), \quad x^1 + i x^2 = r e^{i \phi} .
\end{equation}
Again, the limit in Eq.~\eqref{eq:energyflow2} should be taken while holding $t - r$ constant.

One should be careful about the relative order of the $x^+ \rightarrow \infty$ limit and the $x^- \rightarrow \pm \infty$ integration limits in Eq.~\eqref{eq:energyflow2}. In a gapless theory, everything flows to future null infinity and the order does not matter. But in a gapped theory, as we will explore in the next section, energy is deposited at timelike infinity and the order does matter: one should first integrate over $x^-$ and then take $x^+ \rightarrow \infty$.

We can perform a conformal transformation to relate the energy flow operators to operators defined on the plane $x^+ = 0$. This avoids having to deal with the large $r$ limit. Following~\cite{Hofman:2008ar,Belin:2020lsr}, we consider the transformation
\begin{equation}\label{eq:transformation4D}
    x^+ \rightarrow -\frac{\ell^2}{x^+}, \quad x^{-} \rightarrow x^- - \frac{\abs{x^\perp}^2}{x^+}, \quad x^\perp \rightarrow \ell \frac{x^\perp}{x^+} ,
\end{equation}
where $\ell$ is an arbitrary length scale required to fix the units. Under this transformation the metric transforms as $ds^2 \rightarrow ds^2 (\ell/x^+)^2$. The coordinates in Eq.~\eqref{eq:coordinates} are mapped (in the large $r$ limit) to
\begin{equation}\label{eq:coordinates2}
    x^+ \rightarrow 0, \quad x^- \rightarrow \frac{2(t-r)}{1+\cos \theta}, \quad x^1 + i x^2 \rightarrow \ell e^{i\phi} \tan \left( \theta / 2 \right) .
\end{equation}
In this way the conformal transformation maps the celestial sphere to the $(x^1, x^2)$ plane at $x^+ = 0$. The energy flow operator can be computed on this plane as
\begin{equation}\label{eq:energyflow3}
    \mathcal{E}(\vec{n}) = \left(1 + \abs{x^\perp}^2 \right)^{3} \int_{-\infty}^{+\infty} dx^- T_{--} (x^+ = 0, x^-, x^\perp) ,
\end{equation}
where we set $\ell = 1$ and the $x^\perp$ coordinates are determined from $\vec{n}$ by Eq.~\eqref{eq:coordinates2}. Two of the factors of $1 + \abs{x^\perp}^2$ arise from the transformation of the measure on the celestial sphere, while the third factor comes from the transformation of the $\int dx^- T_{--}$ operator~\cite{Hofman:2008ar}.

\subsection{Shockwave geometries and energy correlators}

We now turn to the computation of energy correlators in theories with a holographic dual. Consider the AdS metric (in lightcone coordinates)
\begin{equation}
    ds^2_{\rm AdS} = \frac{R^2}{z^2} \left( dx^+ dx^- - \left(dx^1 \right)^2 - \left(dx^2 \right)^2 - dz^2 \right)
\end{equation}
where $z \in (0, \infty)$ (with $z = 0$ corresponding to the AdS boundary) and $R$ is the AdS curvature. In what follows we set $R = 1$ --- in other words, we are measuring distances in units of $R$. We assume the 5D weakly-coupled gravitational theory to be dual to a strongly-coupled 4D CFT.

The coordinate transformation in Eq.~\eqref{eq:transformation4D} generalizes as
\begin{equation}\label{eq:transformation5D}
    z \rightarrow \frac{z}{x^+}, \quad x^+ \rightarrow -\frac{1}{x^+}, \quad x^- \rightarrow x^- -\frac{\abs{x^\perp}^2 + z^2}{x^+}, \quad x^\perp \rightarrow \frac{x^\perp}{x^+} ,
\end{equation}
where we have set $\ell=1$.
One can check that this transformation is an isometry, reflecting the fact that AdS space has a built-in conformal invariance.

To study energy correlators, we consider perturbing the CFT action by~\cite{Belin:2020lsr}
\begin{equation}\label{eq:insertion}
    \delta S_{\rm CFT} = \epsilon \int dx^- T_{--} (x^+ = 0, x^-, x^\perp = y^\perp) = \epsilon \int d^4 x T_{--} \delta(x^+) \delta^2 (x^\perp - y^\perp) .
\end{equation}
This effectively inserts the exponentiation of the energy flow operator, Eq.~\eqref{eq:energyflow3}, in the CFT path integral (up to a factor of $(1+ \abs{x^\perp}^2)^3$). Recall that the location $y^\perp$ on the transverse plane corresponds to the particular point on the celestial sphere where we place our calorimeter.

In the dual theory, the insertion of $T_{--}$ acts as a source for the $++$ component of the graviton. Surprisingly, one can find an exact solution of the Einstein equations that corresponds to a localized source on the boundary in terms of a shockwave: a geometry localized in the $x^+$ coordinate that spreads out only in the transverse $x^\perp$ direction as it enters the bulk of AdS ($z>0$). Such a shockwave geometry corresponds to a metric perturbation of the form
\begin{equation}\label{eq:shockwave}
    ds^2 = ds_{\rm AdS}^2 + \frac{\epsilon }{z^2} \delta(x^+) f(x^\perp - y^\perp, z) \left (dx^+ \right)^2 .
\end{equation}
The function $f$ should satisfy the boundary condition $f(x^\perp, 0) \sim \delta^2(x^\perp)$ (the normalization is unimportant for us), reflecting the localization of the $T_{--}$ insertion in Eq.~\eqref{eq:insertion}.

Remarkably, the full Einstein equations for the shockwave ansatz in Eq.~\eqref{eq:shockwave} reduce to a simple linear equation for $f$:
\begin{equation}
    \frac{3}{z} \partial_z f - \left( \partial_1^2 + \partial_2^2 + \partial_z^2 \right) f = 0 .
\end{equation}
The solution which satisfies the correct boundary conditions is well-known and takes the form
\begin{equation}\label{eq:adssoln}
    f(x^\perp, z) = \frac{z^4}{\left(z^2 + \abs{x^\perp}^2 \right)^3} .
\end{equation}
We emphasize that the shockwave is an \textit{exact} solution to the Einstein equations. Moreover, since the equation of motion for $f$ is linear, one can superpose shockwaves to study insertions of multiple energy flow operators at different points:
\begin{equation}
    ds^2 = ds_{\rm AdS}^2 + \frac{\delta(x^+)}{z^2} \left[ \epsilon_1 f(x^\perp - y_1^\perp, z) + \epsilon_2 f(x^\perp - y_2^\perp, z) \right] \left (dx^+ \right)^2 .
\end{equation}
Again, this is an exact solution to the field equations.
One can then take derivatives of the path integral with respect to $\epsilon_{1,2}$ to compute correlation functions.

\subsection{Computing correlators in AdS/CFT}

Once we have the shockwave solution, it is easy to compute energy correlators in the dual theory~\cite{Hofman:2008ar,Belin:2020lsr}. For concreteness and simplicity, let us focus on the case of a scalar external perturbation characterized by momentum $q^\mu = (q, \vec{0})$ (corresponding to energy $q$ and vanishing 3-momentum, as it would be for an $e^+e^-$ collider experiment in the center-of-momentum (COM) frame). This external scalar perturbation will correspond to exciting a bulk scalar field $\phi$ on the AdS side. The important point is that one is now solving for the profile of the bulk scalar in the presence of the shockwave. Since the shockwave is localized in $x^+$ (proportional to $\delta (x^+)$), there will be a discontinuity in the bulk scalar at the shockwave. The scalar equation of motion is
\begin{equation}
    \partial_- \partial_+ \phi + \epsilon \delta(x^+) f(x^\perp, z) \partial_-^2 \phi + \rm{terms~regular~at~shockwave} = 0,
\end{equation}
where we focus only on the terms involving the discontinuity at $x^+ = 0$.
Integrating across the delta function, we can calculate the discontinuity:    
\begin{equation}
    \lim_{\delta \rightarrow 0} \partial_- \phi(x^+ = \delta, x^-, x^\perp, z) =  e^{-\epsilon f(x^\perp, z) \partial_-} \partial_- \phi(x^+ = -\delta, x^-, x^\perp, z) .
\end{equation}
The expectation value of the exponentiated energy flow operator on the state created by $\phi$ then follows as
\begin{equation}\label{eq:expenergyflow}
    \langle e^{\epsilon \mathcal{E}(y^\perp)} \rangle \sim \int \frac{dz}{z^3} d^2 x^\perp d x^- i \phi^* \exp \left[ -\epsilon \left(1 + (y^\perp)^2 \right)^3 f(x^\perp - y^\perp, z) \partial_- \right] \partial_- \phi \Big |_{x^+ = 0} + \rm{~ c.c.}
\end{equation}
Note this is to be evaluated on the shockwave at $x^+ = 0$. In principle we also get contributions away from $x^+ = 0$, but these are irrelevant to the energy flow.

In the absence of the shockwave, the metric is just AdS and the usual AdS propagator determines the scalar wavefunction (together with the boundary value of $\phi$, which is assumed to be a plane wave $e^{i q t}$).
In Appendix~\ref{app:wavefunction} we derive the wavefunction using the embedding of 5D AdS in 6D pseudo-Euclidean space. We show that the scalar wavefunction behaves on the $x^+ = 0$ plane as
\begin{equation}\label{eq:wavefunction}
    \phi(x^+ = 0, x^-, x^\perp, z) \sim \delta(z - 1) \delta^2 (x^\perp) e^{i q x^-/2} .
\end{equation}
We substitute this wavefunction into Eq.~\eqref{eq:expenergyflow} and expand to leading order in $\epsilon$ using the expression for $f$ in Eq.~\eqref{eq:adssoln}, which yields the one-point function up to normalization:
\begin{equation}
    \langle \mathcal{E} \rangle \sim 1 .
\end{equation}
There is no dependence on $x^\perp$ and thus no dependence on the location of the calorimeter on the celestial sphere. To properly keep track of the normalization, we would divide by the norm of the state, which is just Eq.~\eqref{eq:expenergyflow} without the insertion of the shockwave (i.e. when $\epsilon = 0$). We would then obtain $\langle \mathcal{E} \rangle = q/4\pi$, which reproduces the energy of the scalar perturbation upon integration over the celestial sphere~\cite{Hofman:2008ar}. For our purposes we are only interested in the angular dependence.

The two-point function can be computed similarly. We consider inserting two energy flow operators as
\begin{equation}\begin{split}
    \langle e^{\epsilon_1 \mathcal{E}(y_1^\perp)} e^{\epsilon_2 \mathcal{E}(y_2^\perp)} \rangle &\sim \int \frac{dz}{z^3} d^2 x^\perp d x^- i \phi^*  \exp \left[ -\epsilon_1 \left(1 + (y_1^\perp)^2 \right)^3 f(x^\perp - y_1^\perp, z) \partial_- \right] \\
    &\times \exp \left[-\epsilon_2 \left(1 + (y_2^\perp)^2 \right)^3 f(x^\perp - y_2^\perp, z) \partial_- \right] \partial_- \phi + \rm{~ c.c.}
\end{split}\end{equation}
Expanding at leading order in $\epsilon_{1,2}$, we again find a trivial angular dependence, $\langle \mathcal{E} \mathcal{E} \rangle \sim 1$. This corresponds to a spherically symmetric ``mush'' of energy deposition, which is what one would expect at strong coupling.

The essential point is that once one has the shockwave geometry, it is straightforward to compute holographic energy correlators. The correlators are determined by the dependence of the shockwave on $x^\perp / z$. With the full AdS space the results agree with the strongly-coupled CFT expectation.

%%%%%%%%%%%%%%%%%%%%%%%%%%%%%%%%%%%%%%%%%%%%%%%%%%%%%%%%%%%
%%%%%%%%%%%%%%%%%%%%%%%%%%%%%%%%%%%%%%%%%%%%%%%%%%%%%%%%%%%
\section{Energy correlators with an IR brane}\label{sec:IRbrane}
%%%%%%%%%%%%%%%%%%%%%%%%%%%%%%%%%%%%%%%%%%%%%%%%%%%%%%%%%%%
%%%%%%%%%%%%%%%%%%%%%%%%%%%%%%%%%%%%%%%%%%%%%%%%%%%%%%%%%%%

We are now ready to present our calculation of energy correlators for the simplest holographic model of QCD.
For this we cut off the AdS space with an IR brane at $z = z_{\rm IR}$, which serves as a minimal ``hard-wall'' holographic model of confinement. This is essentially a Randall--Sundrum (RSI) model~\cite{Randall:1999ee} with the UV brane sent to the AdS boundary.

This theory is dual to a strongly-coupled CFT which confines at a scale $\sim 1/z_{\rm IR}$. Therefore, if we compute the two-point energy correlator in this model, we should expect to see two qualitatively different regimes. At scales corresponding to energies above the confinement scale, the two-point correlator should be constant, matching the behavior of a strongly-coupled CFT. At energy scales below the confinement scale, the two-point correlator should approach zero, as one would expect for confined hadrons.

We remark that the conformal transformation in Eq.~\eqref{eq:transformation5D}, which is an essential ingredient in the holographic energy correlator calculation, is not an isometry when we introduce the IR brane. However, we can restore the symmetry by imposing that the brane location should transform as $\zir \rightarrow \zir / x^+$. The intuition behind this prescription is that the brane introduces a scale into the theory (namely $\zir$), and so to realize the conformal symmetry we must require that $\zir$ transforms covariantly under conformal transformations.

Since the presence of the brane affects the form of the shockwaves, one might worry that we also need to keep track of how the coordinate transformation affects the shockwaves. We will see that Eq.~\eqref{eq:transformation5D} only rescales the shockwave by an unimportant constant and does not affect its functional form.

\subsection{Calculation of the two-point correlator}

Just as before, we consider shockwave geometries about the AdS metric:
\begin{equation}
    ds^2 = ds_{\rm AdS}^2 + \frac{\epsilon}{z^2} \delta(x^+) f(x^\perp, z) \left (dx^+ \right)^2 ,
\end{equation}
where the shockwave satisfies the boundary condition $f(x^\perp, 0) \sim \delta^2 (x^\perp)$.
The Einstein equations take the same form as in the pure AdS case,
\begin{equation}\label{eq:eomraw}
    \frac{3}{z} \partial_z f - \left( \partial_1^2 + \partial_2^2 + \partial_z^2 \right) f = 0 .
\end{equation}
The crucial difference introduced by the IR brane is to modify the boundary conditions on the shockwave $f$. With the full AdS space, we require that $f$ is regular as $z \rightarrow \infty$. Instead, in RSI the correct boundary condition is a Neumann boundary condition on $f$ at the IR brane~\cite{Randall:1999ee}:
\begin{equation}\label{eq:irbc}
    \partial_z f(x^\perp, z) \Big |_{z = z_{\rm IR}} = 0 .
\end{equation}

To solve the equation of motion, we rescale by a factor of $z^{3/2}$ and Fourier transform over $x^\perp$, defining \begin{equation}
    g(z, k^1, k^2 ) = z^{-3/2} \int dx^\perp e^{-i (k^1 x^1 + k^2 x^2)} f(x^\perp, z) .
\end{equation}
Eq.~\eqref{eq:eomraw} then takes on a Schr\"odinger-like form,
\begin{equation}\label{eq:eomnice}
    \left( k^2 + \frac{15}{4z^2}  - \partial_z^2 \right ) g(z, \vec{k}) = 0 ,
\end{equation}
where $\vec{k} = (k^1, k^2)$ and $k = \abs{\vec{k}}$. The general solution to this equation is
\begin{equation}
    g(z, \vec{k}) = \sqrt{z} \left[ g_1(\vec{k}) K_2( k z) + g_2(\vec{k}) I_2( k z) \right]
\end{equation}
where $g_{1,2}$ are arbitrary functions of $\vec{k}$, to be determined by the boundary conditions.

In terms of $g$, the boundary conditions take the form
\begin{equation}\label{eq:bcnice}
    g(z = 0, \vec{k}) \sim \frac{1}{z^{3/2}} \ (z \rightarrow 0) , \quad  \partial_z \left ( z^{3/2} g(z, \vec{k}) \right ) \Big |_{z = z_{\rm IR}} = 0.
\end{equation}
The UV boundary condition gives $g_1(\vec{k}) \propto k^2$, while the IR boundary condition determines $g_2$. We find
\begin{equation}
    g(z, \vec{k}) \propto k^2 \sqrt{z} \left[ K_2( k z) + \frac{K_1(k \zir)}{I_1(k \zir)} I_2( k z) \right] .
\end{equation}

Finally, we Fourier transform back to position space, yielding a simple expression for our original shockwave $f$:
\begin{equation}\label{eq:shockwavefinal}
    f(x^\perp, z) = \frac{1}{8} \int_0^\infty dk\,  J_0(k r) k^3 z^2 \left[ K_2( k z) + \frac{K_1(k \zir)}{I_1(k \zir)} I_2( k z) \right] ,
\end{equation}
where $r = \abs{x^\perp}$. We have chosen the overall normalization for future convenience. Note that in the absence of the IR brane, only the first term in the square brackets would be present, since requiring regularity at $z \rightarrow \infty$ prohibits the exponentially growing $I_2$ solution. In this case, we can explicitly perform the integration over $k$, which reproduces the usual result in the full AdS space, $f(z, r) = z^4 / (z^2+r^2)^3$. 

One may wonder in which coordinate system the boundary condition Eq.~\eqref{eq:irbc} (as well as the solution Eq.~\eqref{eq:shockwavefinal}) is applicable --- the original physical coordinates, or the ones transformed according to Eq.~\eqref{eq:transformation5D}.  In the original coordinates the IR brane is straight at $z=\zir$, while in the transformed system the brane is located at $z = \zir x^+$. However, to obtain a simple expression for the energy correlators we would like to use Eq.~\eqref{eq:shockwavefinal} for the shockwaves in the transformed frame. This turns out to not be a problem: if we transform the shockwave via Eq.~\eqref{eq:transformation5D}, it only picks up an overall factor of $(x^+)^2$. Since the shockwave is localized in $x^+$, this simply corresponds to a change of its overall normalization, which we have anyway left as arbitrary. Equivalently we may just as well have calculated the shockwave directly in the transformed frame and imposed the BC at $\zir x^+$, since $x^+$ is fixed for a given shockwave. Finally, since the equations of motion for the shockwaves are linear, one can still superpose them without any additional complications.

Once we have the form of the shockwaves in the geometry with the IR cutoff, we can follow exactly the same steps as in Section~\ref{sec:review} to derive the two-point energy correlator. Again taking a scalar source $\phi$, we have
\begin{equation}\label{eq:twopointgeneral}
    \langle \mathcal{E}(0) \mathcal{E}(y^\perp) \rangle \sim \lim_{x^+ \rightarrow 0} \int \frac{dz}{z^3} d^2 x^\perp d x^- i \phi^* f(x^\perp, z)  \left(1 + (y^\perp)^2 \right)^3 f(x^\perp - y^\perp, z)  \partial_-^3 \phi .
\end{equation}
Recall that the $x^+ \rightarrow 0$ limit corresponds to taking the calorimeter to the boundary of Minkowski space, following the coordinate transformation of Eq.~\eqref{eq:transformation5D}. Also, the upper limit of the $z$ integral is $x^+ z_{\rm IR}$ instead of $\infty$ because the extra dimension is cut off by the IR brane. As anticipated below Eq.~\eqref{eq:coordinates}, we will see that for the case with the IR cutoff it is essential to be careful about the relative order of the $x^+ \rightarrow 0$ limit and the integration.

In principle one would need to compute the wavefunction for the source in the geometry with an IR brane, which would no longer take the simple delta-function form in Eq.~\eqref{eq:wavefunction} and make the calculation of the integral very difficult. Fortunately, for energies much larger than the scale of the IR brane ($q \gg 1/z_{\rm IR}$) the source is insensitive to the presence of the brane. Taking $q\gg 1/z_{IR}$ is in fact the right physical regime for a realistic collider like the LHC, where $q$ would be the COM energy $\sim 10$ TeV, while $z_{IR}$ is given by the scale of confinement, $z_{IR}^{-1} \sim$ GeV. It is straightforward to verify that the wavefunction approaches a delta function for large $q$,
\begin{equation}
    \phi \sim \delta(z - \sqrt{1 + x^+ x^-}) \delta^2 (x^\perp) e^{i q (z-1)/x^+} .
    \label{eq:scalarsupport}
\end{equation}
Taking the limit $x^+ \rightarrow 0$ we recover Eq.~\eqref{eq:wavefunction}, as we expect.

\begin{figure}
    \centering
    \includegraphics[width=0.3\textwidth]{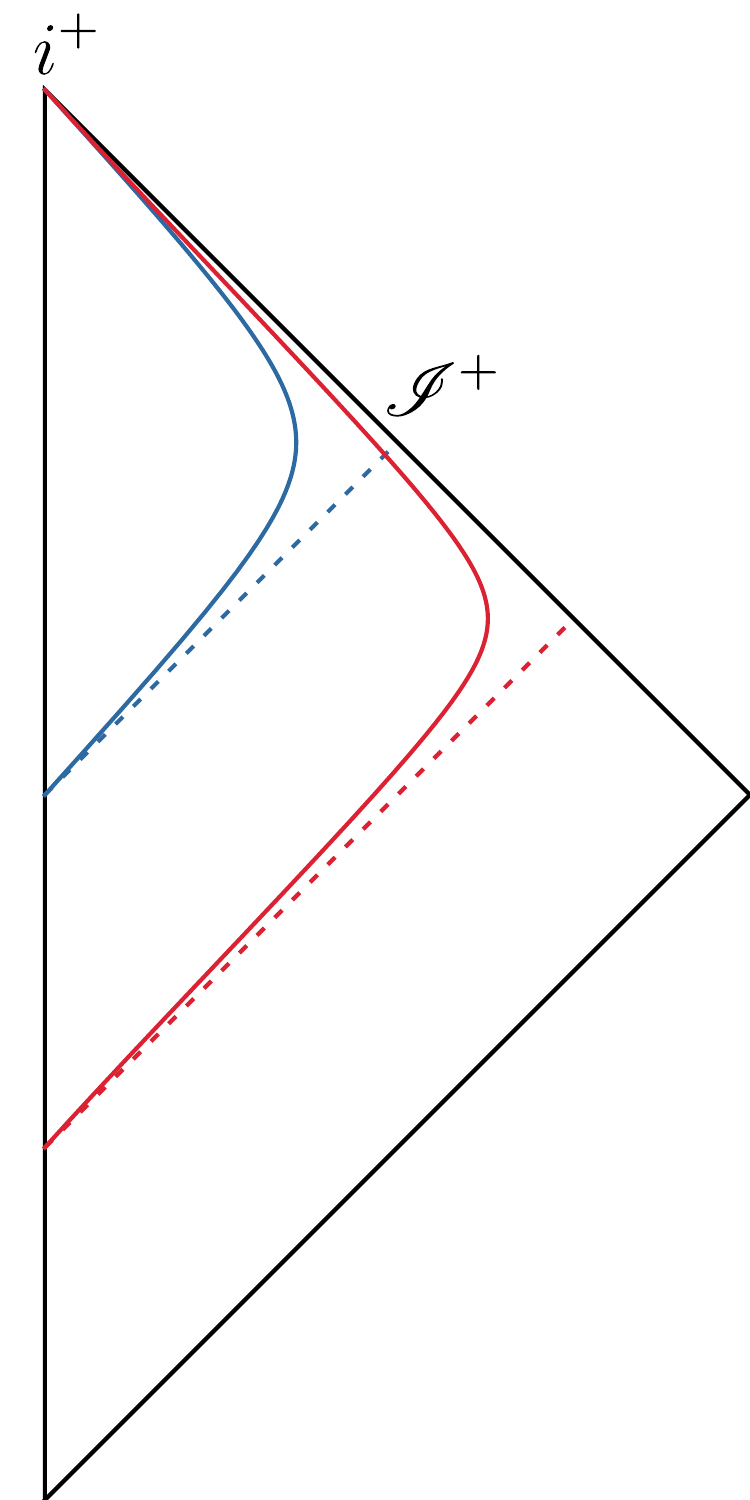}
    \caption{Penrose diagram of the energy flow. In a gapless theory (dashed lines), particles can flow to anywhere on the future null boundary of Minkowski space (denoted by $\mathscr{I}^+$). But in a gapped theory (solid lines), particles end at a single point, future timelike infinity ($i^+$).}
    \label{fig:penrose}
\end{figure}

If we were to na\"ively substitute the wavefunction when the calorimeter is already moved to the boundary of Minkowski space (Eq.~\eqref{eq:wavefunction}) into Eq.~\eqref{eq:twopointgeneral}, it would appear that the $\delta(z - 1)$ lacks support over the integration region when $x^+ < z_{\rm IR}$. Thus as we take the $x^+ \rightarrow 0$ limit the correlator would seemingly vanish.
However this is just an artifact of prematurely sending the calorimeters to the boundary of Minkowski space before performing the integration, essentially taking limits in the wrong order. At nonzero $x^+$ the wavefunction has support over the integration region when $x^- \in (-1/x^+,-1/x^+ + x^+ z_{\rm IR}^2)$ (simply requiring that the delta function be localized between $z = 0$ and $z = x^+ z_{\rm IR}$). As we take $x^+ \rightarrow 0$, there remains support only at a single point in $x^-$. This agrees with the expectation that our theory has a mass gap, so energy flows to future timelike infinity --- a single point on the Penrose diagram, as opposed to future null infinity, where massless radiation flows to (see Fig.~\ref{fig:penrose}). One might then worry that the correlator vanishes in the $x^+ \rightarrow 0$ limit because the range of the $x^-$ integral approaches zero, but we will argue shortly that it remains nonvanishing once it is properly normalized.

Having established the region where the wavefunction has support using Eq.~\eqref{eq:scalarsupport}, the integration over $x^\perp$ and over $z$ is trivial to perform, and we find
\begin{equation}\begin{split}
    \langle \mathcal{E}(0) \mathcal{E}(y^\perp) \rangle \sim \lim_{x^+ \rightarrow 0} &\int_{-1/x^+}^{-1/x^+ + x^+ z_{\rm IR}^2} \frac{dx^-}{z^3} f(0, z)  \left(1 + (y^\perp)^2 \right)^3 f(y^\perp, z) \\
    &\times e^{-iq(z-1)/x^+} \partial_-^3 \left( e^{-iq(z-1)/x^+} \right) \Big |_{z = \sqrt{1+x^+ x^-}} .
\end{split}\end{equation}
The integral over $x^-$ is complicated at nonzero $x^+$ since the integrand depends on $x^-$ (due to the condition $z = \sqrt{1 + x^+ x^-}$). But when we go to the boundary of Minkowski space by taking the $x^+ \rightarrow 0$ limit, the integrand becomes independent of $x^-$. Hence the $x^-$ integral just gives a factor of the integration range $x^+ z_{\rm IR}^2$. This factor cancels out when we properly normalize the correlator (recall the normalization factor is just Eq.~\eqref{eq:twopointgeneral} without the shockwaves inserted), so the correlator does not vanish in the $x^+ \rightarrow 0$ limit. We finally obtain 
\begin{equation}\label{eq:twopoint}\begin{split}
    \langle \mathcal{E}(0) \mathcal{E}(y^\perp) \rangle &\sim \left(1 + (y^\perp)^2 \right)^3 f(y^\perp, z=1)  \\
    &= 1 + \frac{\left(1 + r^2 \right)^3}{8} \int_0^\infty dk J_0(k r) k^3 \frac{K_1(k \zir)}{I_1(k \zir) } I_2(k)  ,
\end{split}\end{equation}
where we used Eq.~\eqref{eq:shockwavefinal} for the shockwave. Recall that we set $R = 1$, so the argument of the $I_2(k)$ is really the dimensionless quantity $k R$. Due to the axial symmetry, the correlator depends only the magnitude of $x^\perp$, not the direction. On the celestial sphere, this corresponds to the correlator only depending on the polar angle $\theta$.

\subsection{Results}

\begin{figure}
    \centering
    \includegraphics[width=0.8\textwidth]{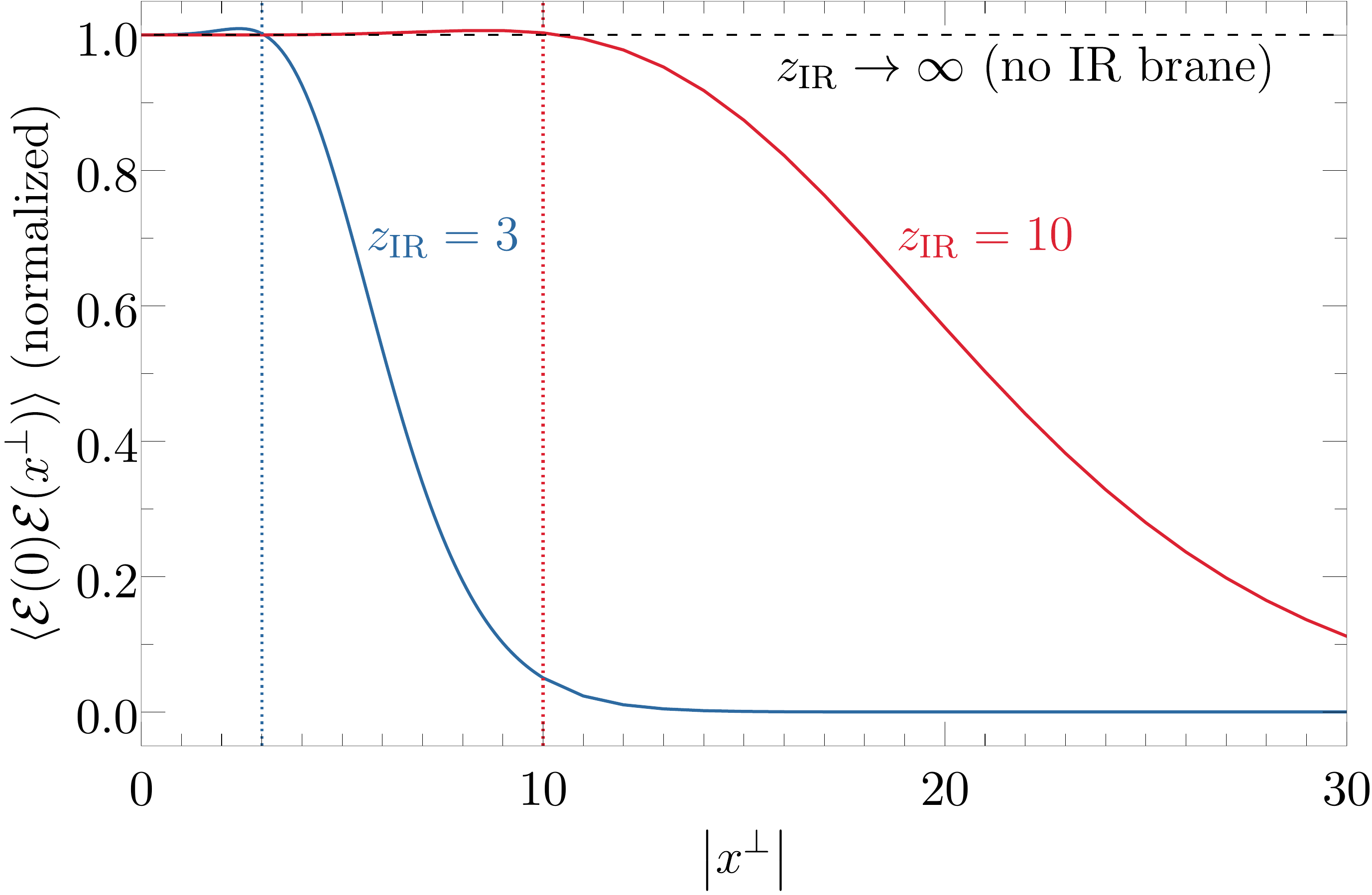}
    \caption{The normalized two-point energy correlator as a function of distance on the transverse plane $\abs{x^\perp}$. We normalize the correlator so that it is equal to one in the absence of the IR brane~\textit{(dashed~black~line)}. We present results for two different locations of the IR brane, $\zir = 3$~\textit{(blue)} and $\zir = 10$~\textit{(red)}. The dotted lines indicate the point at which $x^\perp = \zir$.}
    \label{fig:results}
\end{figure}

In Fig.~\ref{fig:results} we present our main result: the two-point correlator as a function of the transverse separation $r = \abs{x^\perp}$ for two choices of IR brane location $\zir$, computed numerically using Eq.~\eqref{eq:twopoint}. With our choice of normalization, the two-point correlator is one in the $z_{\rm IR} \rightarrow \infty$ limit. Our result agrees with the interpretation we anticipated. For $r \ll \zir$ we probe energies higher than the confinement scale, so the two-point correlator is constant, characteristic of a strongly-coupled CFT. Meanwhile, the correlator falls off for $r \gg \zir$ (we will shortly see that it in fact decays exponentially). We postpone a discussion of the mapping from $r$ to angular separation on the celestial sphere until Section~\ref{sec:discussion}.

\begin{figure}
    \centering
    \includegraphics[width=0.8\textwidth]{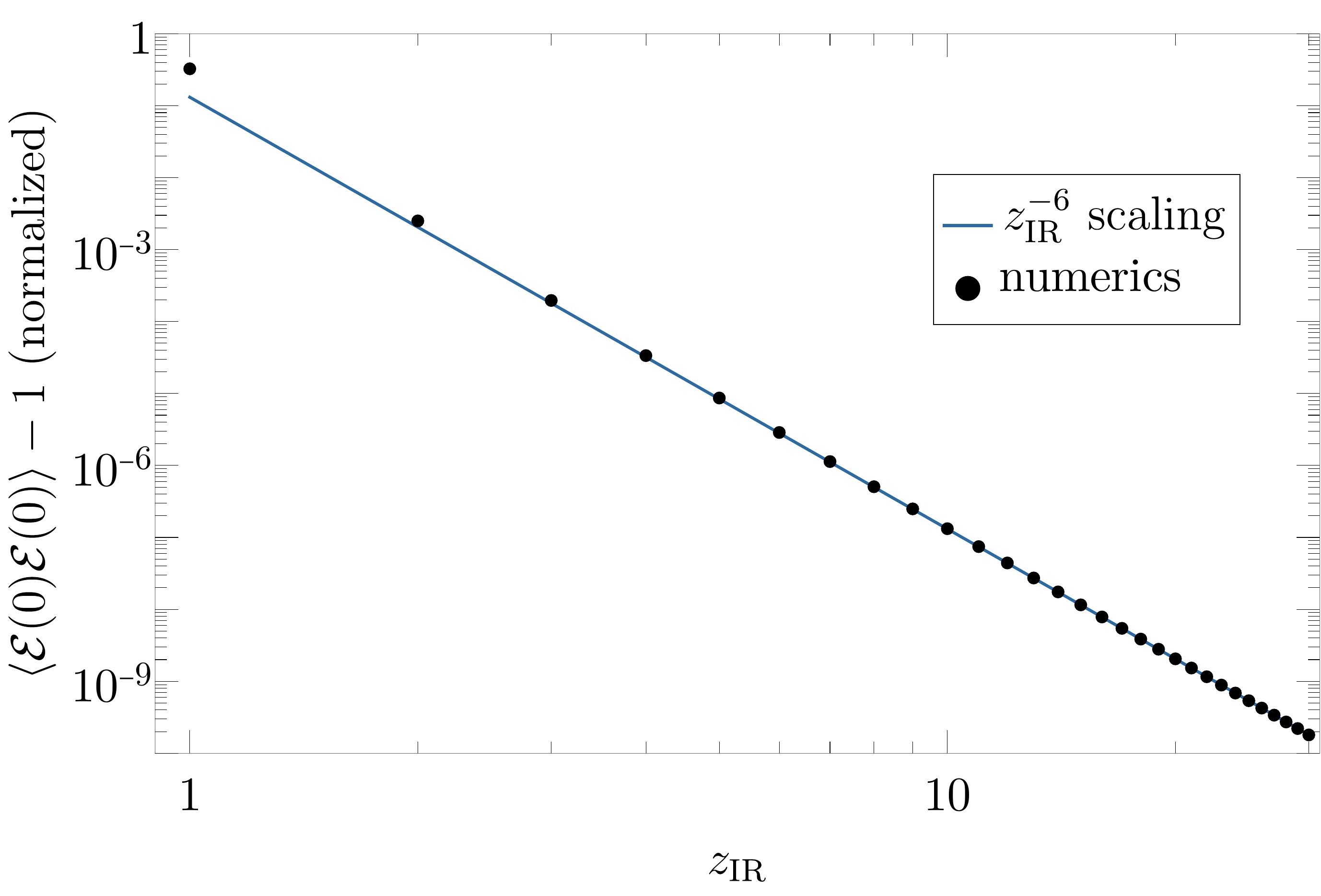}
    \caption{The two-point energy correlator at zero separation as a function of the IR brane location $\zir$. The correlator is normalized as in Fig.~\ref{fig:results} so that $\mathcal{E}(0)\mathcal{E}(0) - 1$ approaches zero for large $\zir$. The black dots show a numerical calculation of the correlator, while the blue line is the $\zir^{-6}$ scaling predicted by Eq.~\eqref{eq:constantterm}.}
    \label{fig:smallr}
\end{figure}

Although the integral in Eq.~\eqref{eq:twopoint} cannot be evaluated analytically, we can still extract the behavior at large and small $r$ to show it agrees with our numerical calculations. Switching variables to $u = k z_{\rm IR}$, the two-point correlator can be written as
\begin{equation}\label{eq:smallr}\begin{split}
    \langle \mathcal{E}(0) \mathcal{E}(r) \rangle - 1 &= \frac{\left(1 + r^2 \right)^3}{8 \zir^6} \int_0^\infty du  u^3 \zir^2 I_2(u / \zir) \frac{K_1(u)}{I_1(u)} J_0(u r / \zir)  \\
    &= \frac{1}{8 \zir^6} \int_0^\infty du  u^3 \zir^2 I_2(u / \zir)  \frac{ K_1(u)}{ I_1(u)} \left[ 1 + \mathcal{O}\left( \frac{r^2}{\zir^2} \right) \right] .
\end{split}\end{equation}
In the second line we have taken the $r \ll \zir$ limit. Thus at small $r$ the two-point correlator approaches a $\zir$-dependent constant (which must approach one as $\zir \rightarrow \infty$ and we recover the full AdS space). We can evaluate this constant asymptotically for large $\zir$. The integrand in Eq.~\eqref{eq:smallr} falls off exponentially for $u \gg 1$ because of the asymptotic behavior of the Bessel functions, and thus we can expand the $I_2(u/ \zir)$ factor for $u \ll \zir$:
\begin{equation}\label{eq:largezir}
    I_2(u/\zir) \approx \frac{u^2}{8\zir^2} .
\end{equation}
Then at $r = 0$ the two-point correlator is approximately
\begin{equation}\label{eq:constantterm}
    \langle \mathcal{E}(0) \mathcal{E}(0) \rangle - 1 \approx \frac{1}{64 \zir^6} \int_0^\infty du  u^5 \frac{K_1(u) }{ I_1(u) } \approx \frac{0.13}{\zir^6}  .
\end{equation}
As expected, the limit as $\zir \rightarrow \infty$ is just one. Eq.~\eqref{eq:constantterm} agrees with a numerical computation, as shown in Fig.~\ref{fig:smallr}.

\begin{figure}
    \centering
    \includegraphics[width=0.8\textwidth]{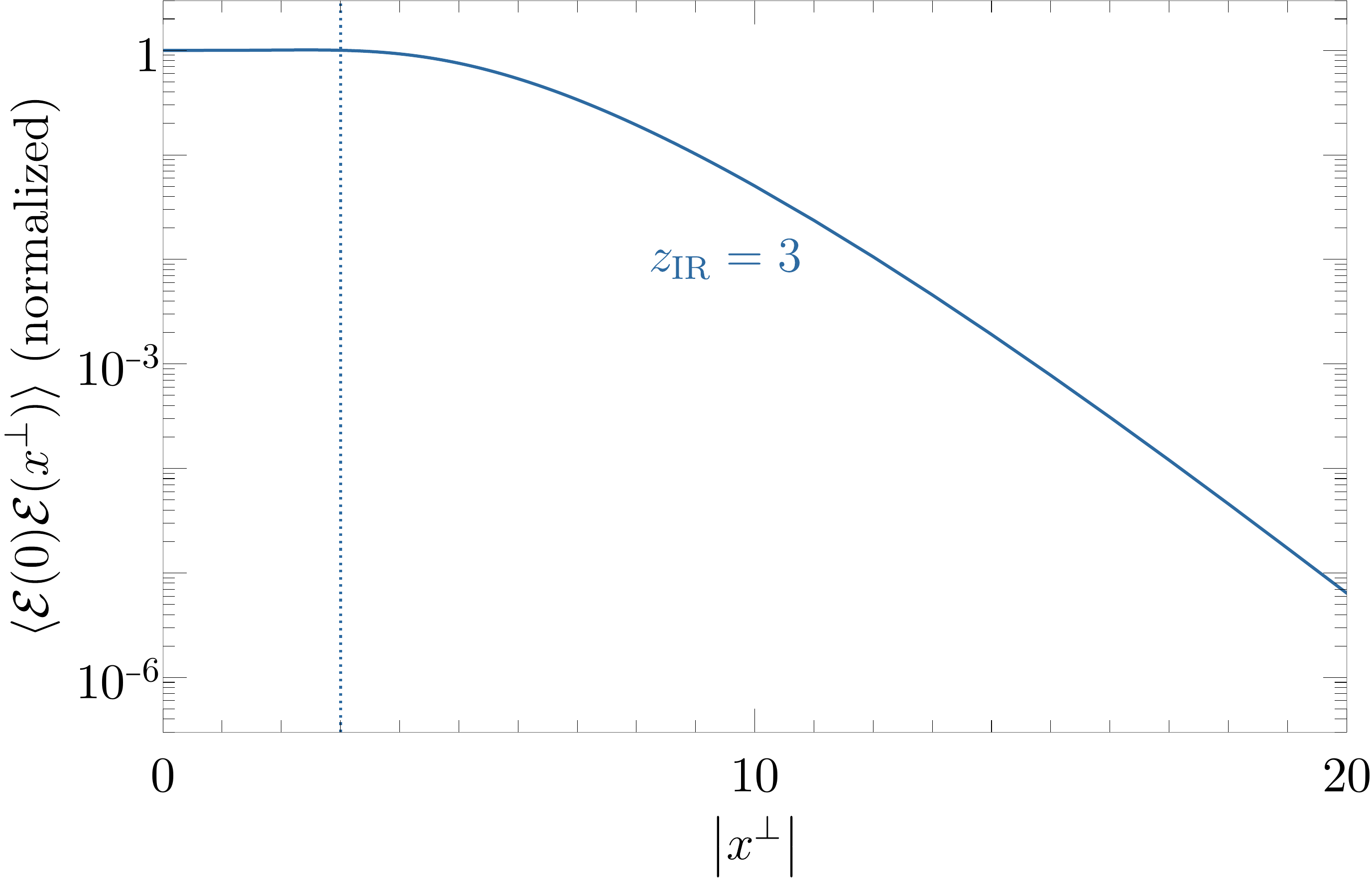}
    \caption{The two-point energy correlator for $\zir = 3$ (same as in Fig.~\ref{fig:results}), but on a log scale. The exponential decay at large distances is manifest.}
    \label{fig:larger}
\end{figure}

Next we discuss the opposite limit, $r \gg \zir$. For simplicity we will again take the limit of large $\zir$ so we can expand the Bessel function as in Eq.~\eqref{eq:largezir}, although our result will be easy to generalize. Then in the $r \gg \zir$ limit we have
\begin{equation}\label{eq:larger}
    \langle \mathcal{E}(0) \mathcal{E}(r) \rangle \sim 1 + \frac{x^6}{64} \int_0^\infty du  u^5 \frac{K_1(u)}{I_1(u)} J_0(x u) ,
\end{equation}
where we defined $x = r/\zir$. The power series expansion of the integrand (not including the $J_0(x u)$ term) begins with
\begin{equation}
    u^5 \frac{K_1(u)}{I_1(u)} \sim 2u^3 + \left(-\frac{3}{4} + \gamma + \log(u/2) \right) u^5 + \mathcal{O}(u^7)
\end{equation}
where $\gamma$ is the Euler--Mascheroni constant.
Importantly, only odd powers of $u$ enter into the series, and the only term involving a logarithm is the $\mathcal{O}(u^5 \log u )$ term written above. A theorem due to Wong~\cite{WongHankel} regarding the asymptotic behavior of Hankel transforms then implies the asymptotic behavior of the correlator is entirely determined by the coefficient of this $u^5 \log u$ term, and we have\footnote{We thank StackExchange user Gary for \href{https://math.stackexchange.com/questions/4847390/asymptotic-behaviour-of-an-oscillating-integral-with-bessel-functions\#comment10327859_4847390}{pointing out this theorem} to us.}
\begin{equation}
    \langle \mathcal{E}(0) \mathcal{E}(r) \rangle \sim 0 .
\end{equation}
Moreover, the correlator must approach the limiting value faster than any power law. This is consistent with the exponential decay seen in our numerical results. In Fig.~\ref{fig:larger} we show the two-point correlator on a log scale, which makes the exponential decay manifest.

%%%%%%%%%%%%%%%%%%%%%%%%%%%%%%%%%%%%%%%%%%%%%%%%%%%%%%%%%%%
%%%%%%%%%%%%%%%%%%%%%%%%%%%%%%%%%%%%%%%%%%%%%%%%%%%%%%%%%%%
\section{Discussion}\label{sec:discussion}
%%%%%%%%%%%%%%%%%%%%%%%%%%%%%%%%%%%%%%%%%%%%%%%%%%%%%%%%%%%
%%%%%%%%%%%%%%%%%%%%%%%%%%%%%%%%%%%%%%%%%%%%%%%%%%%%%%%%%%%

In this work we have presented a holographic computation of energy correlators in a simple model of confinement. Essential to our calculation was the method of calculating shockwave solutions to the Einstein equations. Once we had the form of the shockwave it was trivial to find the two-point correlation function. This method can be used to calculate higher-point correlation functions and should also carry over to other warped backgrounds beyond pure AdS. Given a holographic model with an arbitrary warp factor, one can  repeat our calculations to find energy correlators~\cite{InProgress}.

The two-point correlator revealed a beautiful transition between the constant correlator of a strongly-coupled CFT at high energy scales to an exponential falloff below the confinement scale. Eventually, we would like to reproduce the behavior observed in QCD energy correlators from a holographic model. To this end, let us recapitulate some of the differences between our model and QCD energy correlators, and offer some thoughts on how more sophisticated models can emulate QCD more closely.

The short-distance behavior of our correlator is approximately constant, consistent with the expectation that this model is dual to a strongly-coupled CFT. In QCD, on the other hand, one observes a running characteristic of asymptotically free quarks. Thus, we expect that a more sophisticated AdS/QCD model incorporating asymptotic freedom (such as~\cite{Csaki:2006ji}) would capture this effect~\cite{InProgress}.

A notable feature of our correlator is the exponential decay at large distances. We are not aware of such a feature in QCD energy correlators. It would be interesting to study whether this exponential behavior is universal among gapped holographic theories. The alternative possibility is that it is specific to the hard-wall model of confinement we studied. In this case a different background with a soft-wall type singularity, rather than an IR brane cutoff, could result in different long-distance behavior, such as a power law.

There is another important difference which we glossed over in the previous section. We expressed the correlator as a function of position on the transverse ($x^\perp$) plane, but one can map it back to a position on the celestial sphere via Eq.~\eqref{eq:coordinates2}. One then finds that the low-energy, confined behavior emerges in the back-to-back limit $\pi - \theta \sim 1/\zir$. However, the feature in QCD energy correlators which we originally set out to model appears in the collinear regime~\cite{Komiske:2022enw,CMS:2024mlf}, the opposite limit! Since we are considering a source which is external to the CFT, a more apt point of comparison would be energy correlators of hadrons produced in $e^+ e^-$ collisions, rather than $pp$ collisions as in~\cite{Komiske:2022enw,CMS:2024mlf}. QCD energy correlators in $e^+ e^-$ collisions have been intensely studied, and it is known that hadronization effects are important in both the collinear and back-to-back limits~\cite{Collins:1981va,Collins:1985kw,Fiore:1992sa,Dokshitzer:1999sh,deFlorian:2004mp,Tulipant:2017ybb,Dixon:2019uzg,Ebert:2020sfi}. From this perspective it is expected that we observe effects of confinement in the back-to-back limit; the surprising aspect is that we do \textit{not} observe any effect in the collinear limit. We are unsure why this is the case and it warrants further investigation.

It would be straightforward to extend our results to energy correlators in states produced by vector sources. It is probably more difficult to construct a 5D model of QCD that includes jets, though. In~\cite{Csaki:2008dt} it was argued that incorporating jetty behavior in AdS/QCD requires consideration of stringy dynamics in the 5D gravitational theory.
A complementary perspective is as follows: in a realistic soft-wall model, the curvature blows up as one approaches the soft wall. In this region the 5D gravitational EFT breaks down and quantum gravity will affect the form of the shockwaves. Presumably if one were to resum all the corrections from higher-order gravitational operators in the bulk, the resulting correlators would exhibit jetty behavior.

Our results represent the first step toward a calculation of QCD energy correlators from holographic models. We are optimistic about the future of such computations --- the shockwave approach makes the calculation relatively easy, and we expect that studying more realistic models of confinement will remedy most of the discrepancies between our results and QCD. We hope our work inspires both the communities working on energy correlators and those working on holographic QCD to pursue further research in this area.

%%%%%%%%%%%%%%%%%%%%%%%%%%%%%%%%%%%%%%%%%%%%%%%%%%%%%%%%%%%
%%%%%%%%%%%%%%%%%%%%%%%%%%%%%%%%%%%%%%%%%%%%%%%%%%%%%%%%%%%
%\section*{Acknowledgements}\label{sec:ackn}
\acknowledgments
%%%%%%%%%%%%%%%%%%%%%%%%%%%%%%%%%%%%%%%%%%%%%%%%%%%%%%%%%%%
%%%%%%%%%%%%%%%%%%%%%%%%%%%%%%%%%%%%%%%%%%%%%%%%%%%%%%%%%%%
We are especially grateful to Tom Hartman for several helpful discussions and providing feedback on our manuscript. We also thank David Meltzer, Ian Moult, and Lorenzo Ricci for useful discussions.   CC and AI are supported in part by the NSF grant PHY-2014071. AI is also supported in part by an NSERC PGS D fellowship (funding reference number 557763).   CC is also supported in part by the US-Israeli BSF grant 2016153. CC also thanks the Aspen Center for Physics (supported by the NSF grant PHY-2210452) for its hospitality while this work was initiated. 

\appendix

\section{Scalar wavefunction}\label{app:wavefunction}
In this appendix we derive the scalar wavefunction at the boundary of Minkowski space given in Eq.~\eqref{eq:wavefunction}. It is easiest to do this calculation by embedding 5D AdS in 6D flat space, which we will review following~\cite{Hofman:2008ar}.

\subsection{Embedding of AdS}
We consider a flat 6D space parametrized by coordinates $X^i = \{X^{-1}, X^0, X^1, X^2, X^3, X^4 \}$, with the metric
\begin{equation}
    ds^2 = \left(dX^{-1}\right)^2 + \left( dX^0 \right)^2 - \left( dX^1 \right)^2 - \left( dX^2 \right)^2 - \left( dX^3 \right)^2 - \left( dX^4 \right)^2 .
\end{equation}
In what follows it will be useful to define $X^\pm = X^{-1} \pm X^{4}$. We can describe 5D AdS as a hyperboloid embedded in the 6D space, defined by
\begin{equation}\label{eq:hyperboloid}
    X_i X^i = X^+ X^- + X_\mu X^\mu = R^2
\end{equation}
where $R$ is the AdS curvature scale, which we will set to $1$.

After performing the coordinate transformation in Eq.~\eqref{eq:transformation5D}, the coordinates $x^\mu, z$ parametrize the hyperboloid as
\begin{equation}\label{eq:embedding}
    X^0 + X^3 = \frac{1}{z}, \quad X^\pm = -\frac{x^\pm}{z}, \quad X^\perp = \frac{x^\perp}{z} .
\end{equation}
In the notation of~\cite{Hofman:2008ar}, these are the ``easy coordinates'' denoted by $y^\perp, y^\pm, y_5$. From Eq.~\eqref{eq:hyperboloid} we find
\begin{equation}
    X^0 - X^3 = \frac{1}{z} \left( - x^\mu x_\mu + z^2 \right) .
\end{equation}

\subsection{Calculation of the wavefunction}
We consider a scalar field $\phi$ which propagates in the bulk of AdS and is dual to a CFT operator with scaling dimension $\Delta$. By rewriting the usual AdS propagator in the $X^i$ coordinates, we obtain the following expression for the field configuration~\cite{Hofman:2008ar}:
\begin{equation}\label{eq:propagator}
    \phi(X) = \int d^4 \overline{x} \frac{\phi_0(\overline{x})}{\left( -X^-/2 + X^+/2 (\overline{x}^\mu \overline{x}_\mu) - X^\mu \overline{x}_\mu + i \epsilon \right)^\Delta} .
\end{equation}
Here $\phi_0$ is the value of the field at the AdS boundary in the original Poincar\'e coordinates (before performing the transformation Eq.~\eqref{eq:transformation5D}). We want to consider a scalar source with energy $q$ and no 3-momentum, so we take $\phi_0$ to be the plane wave
\begin{equation}
    \phi_0(x) = e^{i q x^0} .
\end{equation}

It is easiest to evaluate Eq.~\eqref{eq:propagator} for $\Delta = 1$ and then take derivatives with respect to $X^-$ to fix the wavefunction for arbitrary $\Delta$. With $\Delta = 1$ the $\overline{x}^0$ contour integral is straightforward. There are two poles located at $\overline{x}^0 = x_{{\rm pole},\pm}$, where
\begin{equation}
    x_{{\rm pole},\pm} = \frac{X^0}{X^+} \left[1 \pm \sqrt{1 + 2\frac{X^+}{(X^0)^2} \left( \frac{X^-}{2} - \vec{X} \cdot \vec{\overline{x}} + \frac{X^+}{2} \abs{\vec{\overline{x}}}^2\right) } \right]
\end{equation}
After doing the $\overline{x}^0$ integral we are left with
\begin{equation}
    \phi \sim \int d^3 \vec{\overline{x}} \left( e^{i q x_{{\rm pole},+}} + e^{i q x_{{\rm pole},-}} \right) .
\end{equation}
As usual we are not concerned with the overall normalization. To evaluate the remaining integrals one can shift the integration variable to complete the square, such that the integrand only depends on the magnitude of $\vec{\overline{x}}$. This leads to an integral of the form
\begin{equation}
    \int dx e^{i q \sqrt{x^2 + 1/(X^+)^2} } ,
\end{equation}
which can be evaluated in terms of Bessel functions (to do so it is useful to change variables to $t = \sqrt{(X^+ x)^2 + 1}$). We find
\begin{equation}\label{eq:besselwavefunction}
    \phi \sim \frac{e^{i q X^0 / X^+}}{q \left(X^+ \right)^2} J_2 \left( \frac{q}{X^+} \right)
\end{equation}
where $J_2$ is a Bessel function of the first kind.

To recover Eq.~\eqref{eq:wavefunction} we take the limit of Eq.~\eqref{eq:besselwavefunction} as we go to the boundary of Minkowski space at $x^+ = 0$. This corresponds to taking the $X^+ = 0$ limit (see Eq.~\eqref{eq:embedding}). In this limit we have the asymptotic expansion of the Bessel function $J_2(q/X+) \sim \sqrt{X^+/q} e^{-iq/X^+}$, up to an overall constant and phase. We also have $X^0 \sim \sqrt{1 + \abs{\vec{X}}^2 - X^+ X^-}$ using the definition of the hyperboloid, Eq.~\eqref{eq:hyperboloid}. We take $q$ to have a small and positive imaginary part $q \rightarrow q + i \epsilon$, which is just a way to implement the usual $i\epsilon$ prescription. Using these results we obtain
\begin{equation}
    \lim_{X^+ \rightarrow 0} \phi \sim \lim_{X^+ \rightarrow 0} \frac{e^{iq/X^+ (\sqrt{1 + \abs{\vec{X}}^2 - X^+ X^-} - 1)}} {\left(q X^+ \right)^{3/2}} = \frac{e^{-i q X^- / 2}}{q^3} \delta^3 \left( \vec{X} \right) .
\end{equation}
For an arbitrary value of $\Delta$ the $q^3$ factor is replaced by $q^{4-\Delta}$.

Lastly, we convert the delta function back to the $x^\mu, z$ coordinates. At the boundary $X^+ = 0$, the point $\vec{X} = 0$ corresponds to $X^0 = 1$ by the definition of the hyperboloid, Eq.~\eqref{eq:hyperboloid}. Then using Eq.~\eqref{eq:embedding} it is clear that this corresponds to the point $x^\perp = 0$, $z = 1$ in the $x^\mu, z$ coordinates. This leads to Eq.~\eqref{eq:wavefunction},
\begin{equation}
    \phi \sim e^{i q x^- / 2} \delta^2 \left( x^\perp \right) \delta(z - 1) ,
\end{equation}
as desired.

\bibliographystyle{JHEP}
\bibliography{references}

\end{document}